Received 31 October 2025, accepted 6 December 2025, date of publication 17 December 2025, date of current version 23 December 2025.

*Digital Object Identifier 10.1109/ACCESS.2025.3645466*

PERSPECTIVE

# Vibe Coding as a Reconfiguration of Intent Mediation in Software Development: Definition, Implications, and Research Agenda

**CHRISTIAN MESKE[1,2], TOBIAS HERMANNS[1], ESTHER VON DER WEIDEN[1], KAI-UWE LOSER[1], AND THORSTEN BERGER[2]**

[1]Faculty of Mechanical Engineering, Ruhr University Bochum, 44801 Bochum, Germany
[2]Faculty of Computer Science, Ruhr University Bochum, 44801 Bochum, Germany

Corresponding author: Christian Meske (christian.meske@rub.de)

**ABSTRACT** Software development is undergoing a fundamental transformation as vibe coding becomes widespread, with large portions of contemporary codebases now being generated by Artificial Intelligence (AI). The disconnect between rapid adoption and limited conceptual understanding highlights the need for an inquiry into this emerging paradigm. Drawing on an intent perspective and historical analysis, we define vibe coding as a software development paradigm where humans and Generative AI (GenAI) engage in collaborative flow to co-create software artifacts through natural language dialogue, shifting the mediation of developer intent from deterministic instruction to probabilistic inference. By intent mediation, we refer to the fundamental process through which developers translate their conceptual goals into representations that computational systems can execute. Our results show that vibe coding redistributes epistemic labor between humans and machines, shifting expertise from technical implementation toward collaborative orchestration. We identify key opportunities, including democratization, acceleration, and systemic leverage, alongside risks such as black-box codebases, responsibility gaps, and ecosystem bias. We conclude with a research agenda spanning human-, technology-, and organization-centered directions to guide future investigations of this paradigm.

## I. INTRODUCTION

The software development landscape is undergoing a profound transformation. Organizations that fund and mentor early-stage startups in Silicon Valley report that 25% of companies in their Winter 2025 cohort had codebases that were 95% AI-generated [1]. Further, analyses of enterprise codebases indicates that the adoption of AI-assisted development has substantially increased code creation velocity, with pull requests surging by 70% since late 2022, even as the number of developers has remained steady [2]. Such statistics reflect the rapid emergence of what Andrej Karpathy [3] has termed "vibe coding," a conversational way of creating artifacts where developers "see stuff, say stuff, run stuff" in dialogue with Artificial Intelligence (AI) systems, fundamentally altering how software is conceived and created. Rather than crafting an artifact through code line by line, developers are increasingly enabled to articulate higher-level intentionality through open-ended conversational loops, in which AI not only generates code, but also engages in problem framing and iterative sense-making [4]. While vibe coding shares surface similarities with other forms of AI-assisted development, it represents a qualitatively distinct practice as described by Karpathy. Instead of merely using prompts to generate code snippets, vibe coding describes an entire development process conducted through natural language dialogue. The focus, therefore, lies not on prompt engineering, which concerns the deliberate crafting and refinement of inputs to optimize discrete AI outputs, but on an iterative, conversational workflow in which software artifacts emerge through sustained human-AI exchange. Additionally, traditional co-creative coding assistants presuppose a developer who remains immersed in the codebase, working at the level of syntax and implementation, while the AI offers recommendations that are reviewed and selectively integrated. Vibe coding inverts this relationship: the developer operates at the conversational level, guiding through high-level intent and iterative feedback, while the AI assumes responsibility for implementation.

This shift represents more than a technological convenience; it marks a fundamental reconfiguration of intent mediation in software development ,which we understand as the process of translating conceptual goals into representations that computational systems can execute has been central to software development [5], [6]. For instance, Norman [56] refers to the Gulf of Evaluation and the Gulf of Execution as gaps between user intent and system response, while Leveson [6] demonstrated how system purpose and design principles must be systematically translated into executable representations. Throughout computing history, major paradigm shifts have transformed how humans translate desired outcomes into machine-executable instructions: from the physical manipulation of hardware circuits in systems like Electronic Numerical Integrator and Computer (ENIAC) and Zuse's Z3 [7], [8], through symbolic abstractions like assembly and high-level languages such as Fortran and Algol [9], [10], to object-oriented paradigms and integrated development environments [11]. Each transition fundamentally altered the cognitive demands and epistemic requirements of software development. Especially Nygaard [12] was interested in democratizing the development process, making it more accessible to end-users. Vibe coding can be seen in line with this development, but creates a different kind of shift: from deterministic instruction, where developers must explicitly encode intent through formal syntax, to probabilistic interpretation, where AI systems infer meaning from naturalistic expression and assume responsibility for translating human goals into executable code. This transformation extends far beyond productivity gains, fundamentally reshaping who can develop software [13]. Vibe coding reframes software development as interpretive co-creation, where humans and AI agents collaboratively construct solutions through iterative dialogue and mutual interpretation, rather than formal construction, where developers must explicitly design and execute all implementation details through predetermined syntax and logical structures, aligning with broader theories of distributed cognition [14] and hybrid intelligence, where cognitive work is dynamically shared between developer and AI agents [15]. The implications of the transformation span the reconfiguration of individual cognitive work, the evolution of professional expertise, and organizational structures, as software development shifts from a specialized craft requiring years of technical mastery to a more accessible, conversational approach, where domain knowledge and strategic thinking become more important as implementation skills, while simultaneously introducing risks of technical deskilling, responsibility gaps, and code quality concerns that challenge established software engineering practices.

While surveys show substantial integration of AI assistants into everyday workflows [16], [17], the conceptual understanding of vibe coding remains underdeveloped. Existing research has examined the integration of Large Language Models (LLMs) into software engineering tasks, highlighting practical benefits in code generation and productivity gains [17], [18]. But many of these approaches predominantly view LLMs as subordinate assistants within conventional development paradigms rather than as collaborative partners in a fundamentally new way of creating artifacts. For instance, Gao et al. [18] systematize technical architectures and performance metrics across LLM variants, while setting aside questions of how such models reconfigure the nature of programming in software development and intent mediation itself. This disconnect between widespread tool adoption and conceptual understanding reflects a broader pattern where practical use outpaces theoretical frameworks, creating urgent needs for systematic analysis of this emerging paradigm. We address this gap by providing the first systematic conceptualization of vibe coding as a distinct programming mode and analyzing its implications for software development practice. Our research is guided by two questions:

1) How can vibe coding be defined as a distinct software development paradigm, and how does it reconfigure the mediation of developer intent compared to traditional practices?
2) What cognitive, epistemic, and organizational implications, both beneficial and problematic, arise from vibe coding?

To investigate these questions, in Section II we begin by tracing the evolution of intent mediation in software development since the 1940s, identifying structural and epistemic shifts across the epochs. In Section III, we then define vibe coding in contrast to to traditional software development, articulating its key attributes and interaction patterns, which anchors our analysis of how vibe coding reconfigures cognitive work, expertise, and epistemic agency. Drawing on this conceptual groundwork, in Section IV we synthesize opportunities (e.g., accessibility, democratization, acceleration) and risks (e.g., deskilling, opacity, responsibility gaps) that emerge from the interpretive nature of vibe coding. In Section V, we critically reflect on the findings and outline future research directions. This paper ends with a conclusion and discussion of limitations in Section VI.

## II. HISTORY OF INTENT MEDIATION IN SOFTWARE DEVELOPMENT

Intent mediation in software development has evolved significantly over the decades, reflecting changes in both how intent is expressed and how cognitive effort is distributed between human and machine. This section traces that evolution across nine decades, each marked by a significant evolution regarding the form of mediation and the nature of software development work. Each era concludes with a synthesis that reflects on the dominant patterns and implications for how intent was conveyed during that period. Together, these historical developments, summarized in Table 1 at the end of Section II, provide a basis for understanding how the mediation of intent has shaped and continues to shape the practice of software development.

### A. MANUAL TRANSLATION: HARDWARE MANIPULATION TO ALGORITHMIC SPECIFICATION (1940S-1960s)

In the 1940s programmers mediated intent by physically manipulating machine components [19], [20]. On the ENIAC, programs were "constructed" by manually setting switches and connecting patch cables [19], while Zuse's Z3 used punched tape to feed instructions into fixed hardware circuits [7], [21], [22]. Both were tightly coupled to hardware architecture with no separation between logic and machine operation [8], [23]. Each system required its own approach, making programming an inherently machine-specific task [7].

The advent of assembly languages in the 1950s marked the transition from physical manipulation of hardware to using symbolic expressions [19]. Rather than configuring cables or switches, developers were able to use instructions mimicking natural language. Short textual codes, such as 'ADD,' 'MOV,' or 'JMP' were directly mapped to the machine's binary operations [21], [23], [24]. The new layer of abstraction allowed to mediate intent in a language-like form that was more human-readable, easily modifiable, and replicable compared to manual hardware reconfigurations [25]. Despite the use of textual mnemonics, the development process remained closely tied to machine architecture. Each symbolic instruction still corresponded one-to-one with specific hardware actions, still requiring developers to think in terms of memory addresses, Central Processing Unit (CPU) registers, and exact sequencing of low-level operations [26].

From the late 1950s through the 1960s, programming underwent a significant leap in abstraction, moving beyond the one-to-one symbolic mediation of assembly languages. New high-level programming languages like FORTRAN, ALGOL, COBOL and C moved away from hardware-specific encodings [21], [27], instead emphasizing machine-agnostic, higher-order constructs such as loops, conditionals, and functions. These constructs enabled the formal specification of complex procedural logic [21], [28]. Developers could now mediate intent through single high-level statements [21]. For example, a simple loop could be expressed in one concise line in a high-level language, whereas achieving the same in assembly would require manually managing memory addresses, loop counters, conditional jumps, and instruction flow control in multiples lines of code. Compilers provided the required software capabilities that translated these abstract algorithmic statements into the multitude of low-level instructions required for execution on a specific machine [9], [23], [27], [29]. Overall, this era was marked by efforts to formalize the nature and structure of programming languages, defining programming language grammar [10], establishing concepts such as lexical scoping and block-structured control constructs [10], [30], promoting separation of concerns, abstraction boundaries, and systematic decomposition [6].

The foundational era from the 1940s to the 1960s, thus, demonstrates a profound evolution in how developers mediate intent and engage cognitively with computational systems. From the physical manipulation of hardware circuits requiring intimate machine-specific knowledge, through assembly's symbolic mnemonics that maintained one-to-one hardware correspondence, development culminated in high-level languages that enabled abstract algorithmic intent mediation independent of underlying architecture. This progression fundamentally transformed cognitive work from hardware-focused mechanical controlling to conceptual algorithmic thinking, establishing the foundations for programming as an intellectual discipline.

### B. CONCEPTUAL MODELING: STRUCTURED PROGRAMMING TO DESIGN PATTERNS (1970s-1990s)

By the 1970s, structured programming had become dominant, with developers writing procedural logic step by step. This proved increasingly tedious and error-prone [31], [32], motivating declarative languages like Structured Query Language (SQL) and Prolog [33], [34] that shifted focus from defining procedures to specifying conditions [35]. Declarative programming inverted the programmer's relationship with the machine: instead of instructing how to compute a result, one specifies the desired outcome [37]. With SQL, for instance, a developer does not define the procedural steps for accessing and comparing data, instead, they write a single formal statement that describes the structure of the result, leaving the execution strategy to the machine [38]. Similarly, Prolog represented a distinct branch of declarative programming known as logic programming, allowing to define a set of logical facts. Computation then becomes a process of machine-automated resolution: the system searches for results that satisfy a query, automatically applying inference steps that were not explicitly spelled out [39]. In parallel, functional programming offered another alternative to procedural expression of intent. Building on the foundations of early languages like Lisp, functional programming languages such as Scheme and Meta Language formalized computation around the concept of mathematical functions [40], allowing intent to be expressed without stepwise manipulation of state. Like declarative programming, it offered a model where developers could describe what should be computed, while abstracting away from how individual steps were executed [41].

By the 1980s, growing software complexity strained procedural code [21], [23]. Object-oriented programming (OOP) emerged to address this by redefining intent mediation around objects, self-contained entities combining data and behavior [42], [42], allowing developers to model real-world concepts as interacting objects rather than global procedures [19], [27], [43]. Languages like Simula, Smalltalk and later C++ allowed developers to define classes, encapsulate state, and structure programs around message-passing between objects [43], [44]. While intent mediation still occurred through structured programming languages with defined syntax and semantics, developers increasingly approached problems not just through fixed sequences of steps, but by thinking in terms of distinct roles and responsibilities within code. Instead of focusing solely on controlling a singular flow of execution, they began to describe systems in terms of how different parts should interact, offering an alternative mental model to procedural logic.

This shift toward expressing intent through conceptual structures continued into the 1990s with the emergence of design patterns that provided reusable templates that structure software systems and communicate underlying intent consistently [45]. First formalized by Gamma [46], design patterns encapsulate proven solutions to recurring problems encountered in software development. They offer developers a shared vocabulary and a set of best practices that make the underlying design intent more explicit and communicable. Design patterns thus mediate intent not only at the level of individual components but across whole system architectures, embedding requirements and domain logic into reusable forms [47], [48]. By formalizing these solutions, design patterns, by design, mediate the developer's intent, ensuring that underlying principles and requirements are consistently understood and implemented [49].

The 1970s to 1990s, thus, marked a turn from purely procedural control toward more expressive design and mediation of intent. As programming languages and paradigms matured, expressing intent became less about operational detail and more about developers shaping conceptual structures. The cognitive work of programming in software development shifted from managing stepwise execution to articulating coherent designs that reflect how developers understand and frame problems. Instead of translating intent into granular instructions, they began shaping code in forms that aligned with their mental models.

### C. COLLABORATIVE SYNTHESIS: FROM PREDICTIVE ASSISTANCE TO AI CO-CREATION (2000s-2020s)

The 2000s saw low-code and no-code platforms emerge, enabling users without expertise to build systems by selecting templates, configuring modules, and assembling prebuilt components through graphical dashboards [50], [51], [52]. This abstracted away boilerplate code [54], [55], making development a matter of selecting features and orchestrating workflows through predefined options. Logic was embedded in the interface itself, constraining and guiding what could be expressed [51], [54], [56]. This approach did not replace traditional software development but introduced an alternative model of intent mediation, one where assembly and configuration took precedence over manual authoring and abstraction, shifting some of the responsibility to the machine to anticipate and interpret the developer's intent. While low-code systems expanded participation, traditional programming remained dominant with IDEs providing static code completion based on lexical rules. In the 2010s, recognition that code exhibited statistical regularities similar to natural language [57], [58] enabled machine learning support [59]. Code completion evolved from nearest-neighbor models [60] to Bayesian networks [61], moving beyond static suggestions to actively interpreting developer intent in context. By the late 2010s, neural models further advanced this approach, learning to predict more context sensitive completions like fitting variable names [62], [63]. These models offered completions that were not only syntactically valid but semantically plausible. While intent was still mediated through traditional code, the nature of interaction changed. Developers engaged in a new kind of dialogue with their tools, assisted by systems that could anticipate intent, transforming the development process into a more assisted activity.

In the 2020s, large language models integrated directly into developer workflows through GenAI [64], [65], anticipating intent and proposing syntactically correct, contextually relevant code. This mediation takes two primary forms. The first is in-line assistance, exemplified by tools like GitHub Copilot, which extends the concept of autocomplete from a single keyword to entire multi-line function blocks [64], [66]. As the developer types a comment or a function signature, GenAI offers a complete implementation as "ghost text", which can be accepted, rejected, or modified [67]. The second form is conversational snippet generation. Here, the developer might temporarily leave the IDE to engage in a dialogue with a LLM like ChatGPT, e.g., asking it to "write a function that generates prime numbers" or "generate ideas how to build an app for image filteres" [65]. The developer then acts as a curator of ideas and generated code [65].

This symbiotic interaction profoundly reallocates cognitive work. The primary burden is no longer the meticulous authoring of every aspect of a artifacts logic. Instead, it shifts to more high-level tasks like prompt articulation, expert supervision, and careful integration. The developer's core cognitive work becomes formulating a clear request (prompt) [68], critically evaluating the AI's output for correctness, security, and efficiency, and then weaving that generated output into the larger fabric of the application. The result ceases to be a self-authored artifact and becomes a collaborative piece that was co-created with the GenAI. While this level introduces a form of natural language interaction, it remains firmly grounded in the production of discrete, syntactically-bound code snippets, a crucial distinction from the more holistic, goal-oriented mediation that would follow with vibe coding.

The 2000s to 2020s thus marked a decisive shift toward assisted and collaborative forms of intent mediation. Beginning with low-code platforms that allowed users to configure systems through graphical assembly rather than manual coding, intent mediation expanded to include new user groups and interaction models. As statistical patterns in code were recognized and exploited, development environments evolved from passive editors into predictive, context-aware assistants. This trajectory culminated in the integration of LLMs that engage developers in a form of co-creation, where intent is expressed not just through code but through natural language prompts and ongoing dialog. Across these developments, the cognitive work of developers transitioned once again: from authoring and structuring logic manually, to orchestrating, curating, and supervising machine-generated contributions. Software development became a mediated activity not just through tools, but through shared agency between human and machine.

## III. VIBE CODING: FROM DETERMINISTIC TO PROBABILISTIC INTENT MEDIATION

Building on the historical trajectory outlined in Section II, this section introduces vibe coding as a new software development paradigm that reconfigures how intent is mediated and how cognitive work is shared between humans and machines. Unlike traditional approaches, where developers explicitly encode goals into formal structures, vibe coding centers on interpretive collaboration with Generative AI that infers, adapts, and implements intent expressed in natural language. The discussion is divided into two parts. Section III-A defines vibe coding as a conversational, multimodal software development paradigm, marked by co-creative timing and semantic-level abstraction. Section III-B examines how this model reshapes cognitive demands and redistributes development expertise, proposing a new configuration of epistemic agency.

### A. DEFINITION AND CONCEPTUALIZATION OF VIBE CODING

The notion of "vibe" in software practice evolved from informal discourse among developers to describe collaborative rhythm and cognitive alignment in AI-assisted work. When Andrej Karpathy [3] formalized this as "vibe coding" in early 2025, he crystallized what had become a recognizable interactional pattern, one that has quickly gained traction in both developer communities and industry media [3], [69]. In this sense, vibe coding represents the latest stage in a longer evolution from structured programming and pair programming toward conversational, co-creative modes of software production. LLM-powered IDEs such as GitHub Copilot, Amazon CodeWhisperer, Tabnine, and specialized agents (e.g., Replit's Agent, Devin, Claude Code) are central to the vibe coding workflow. Unlike traditional workflows, vibe coding allows users to "see stuff, say stuff, run stuff" [3] in conversational flow [70], prioritizing intuitive expression over technical specification [71] through probabilistic generative programming.

Notably, Y Combinator [1], renowned startup incubator and venture capitalist from Silicon Valley, reported that 25% of startup companies in its Winter 2025 batch had codebases that were 95% AI-generated, reflecting a move toward AI-assisted development. To understand the significance of this paradigm shift, we must first examine the term itself.

"Vibe" in contemporary discourse refers not only to an ambient emotional atmosphere, but also to a state of resonant interaction: Merriam-Webster [72] defines it as "a distinctive feeling or quality capable of being sensed", highlighting its subjective and relational nature. Colloquially, to "vibe" refers to aligning and harmonizing with another entity, where interaction feels effortless and flow emerges naturally. This phenomenon parallels the concept of networked flow, where creative collaboration flourishes when participants experience strong social presence and collective immersion, enabling seamless idea generation and shared understanding [73]. Similarly, effective teamwork depends on establishing harmony and rhythm across learning modes, which enhances cohesion and drives innovation [74]. These insights suggest that "vibing" is more than a colloquialism: it describes a critical collaborative dynamic in which resonance and synchronization foster successful outcomes. This cultural context illuminates why "vibe coding" aptly captures the essence of this new approach: it emphasizes a synchronous co-creative dialogue where developer and AI find a collaborative rhythm, developing solutions through iterative conversation rather than precise technical specification. These transformations mark a transition from deterministic to probabilistic intent mediation. Accordingly, we define vibe coding as follows

> *Vibe coding is a software development paradigm where humans and Generative AI engage in collaborative flow to co-create software artifacts through natural language dialogue, shifting the mediation of developer intent from deterministic instruction to probabilistic inference.*

Vibe coding manifests through five key attributes: (1) goal-oriented intent expression describing goals rather than implementation; (2) rapid dialogic interaction replacing write-compile-test loops with conversational feedback; (3) implementation abstraction enabling deployment without full understanding of details; (4) dynamic semantic refinement where requirements evolve through AI interpretations; and (5) co-creative flow states establishing productive rhythm between developers and AI. Figure 1 represents these attributes through a simplified example, where a user enters a dialogue with a Generative AI with the aim to vibe-code a website.

While traditional development environments passively expected explicit commands, vibe coding introduces AI as an epistemic agent that, in response to human intention and interaction, actively participates in knowledge construction, interpretation of requirements, and collaborative sensemaking. Rather than merely executing predefined instructions, it participates in the interpretation of the developer's explicit or implicit goals based on inference, predicting potential needs, and offering output informed by patterns learned across vast code repositories. This epistemic dimension fundamentally reshapes the developer-machine relationship. The AI becomes capable of contributing solutions that may exceed the developer's technical knowledge, identifying ambiguities in requirements that need clarification, and recommending alternative implementation strategies based on its extensive pattern recognition capabilities. These dynamics set the stage for our next section, which examines concrete examples of vibe coding and explores the resulting changes in cognitive work and technical expertise.

**TABLE 1.** Intent Mediation in Software Development (1940s–2020s).

| Era | Anchor | Form of Intent Mediation | Cognitive Work |
|---|---|---|---|
| 1940s | Hardware Control | Physical manipulation of switches, plugboards, and wires. | Translating logic directly into a physical machine configuration. |
| 1950s | Symbolic Code | Textual mnemonics representing machine opcodes. | Meticulously managing CPU registers and memory addresses. |
| 1960s | High Level Languages | Structured, high-level textual syntax with formal grammar. | Designing step-by-step logic and managing the state of variables. |
| 1970s | Declarative Paradigm | Domain-specific, descriptive statements that define desired results. | Specifying the "what" and delegating the "how." |
| 1980s | Object Oriented Programming | Mapping of real-world entities that embody roles and responsibilities. | Conceptual modelling of real-world entities. |
| 1990s | Design Patterns | Usage of templates that carry underlying intent. | Identifying common problems and applying proven solutions. |
| 2000s | Component Configuration | Assembling and configuring pre-built visual components. | Orchestrating systems by shaping behavior through interface constraints. |
| 2010s | ML Predictive Assistance | Partial single code lines interpreted by ML. | Reviewing, editing, and integrating context-sensitive completions. |
| 2020s | LLM Code Generation | Single code lines, code-contexts, and natural language prompts interpreted by LLMs. | Prompting, reviewing, and correcting contextually generated code snippets. |

Figure 2 illustrates the paradigm shift from deterministic intent mediation in traditional software development to probabilistic intent mediation in vibe coding. In both scenarios, the human actor begins with a specific intent. In our example, the developer carries the intent "I want to sort the list [3, 1, 2] from smallest to largest." (1). To enable computer execution, the developer must first overcome the intent mediation gap between the human and the computational system. Traditionally, this mediation requires the adherence to a rigid and narrow specification space. The developer must produce code that conforms exactly to predetermined syntactic and semantic rules for deterministic, instructional execution (2). In contrast, vibe coding allows developers to bridge this gap through natural language communication, which operates on probabilistic-interpretive principles. The developer mediates intent through interaction with a LLM (3), which assumes responsibility for interpreting the natural language specification and producing executable code outputs (4). Regardless of the mediation pathway, whether through direct deterministic coding (2) or LLM-interpreted probabilistic communication (3), the resulting code output undergoes deterministic processing (e.g., compilation) before execution by the processor, maintaining the same final computational determinism in both paradigms (5).

### B. RECONFIGURING COGNITIVE WORK AND EXPERTISE

Building upon the exploration of vibe coding as a collaborative, natural language, dialogic co-creation flow, this section examines how this emerging approach fundamentally reconfigures cognitive demands and expertise in the software development practice. The foundation begins with "cognitive alignment," where mental models and artifacts emerge through natural language, which enables a new "cognitive work division" as tasks redistribute between human and AI partners. This redistribution naturally alters the "cognitive rhythm" of development work, so we argue, creating more fluid patterns of engagement. These changes in turn reshape the "nature of expertise" from technical mastery toward collaborative orchestration. As expertise evolves, "knowledge boundaries" become more permeable, with AI compensating for human knowledge gaps. Finally, these shifting boundaries necessitate reconsideration of "epistemic agency and responsibility" when artifact creation becomes a shared human-AI endeavor. By examining these dimensions, we reveal how vibe coding creates a fundamentally different cognitive landscape for developers. Table 2 at the end summarizes Section III-A and Section III-B by systematically contrasting traditional programming with vibe coding.

Traditional programming often forces developers to translate high-level goals into low-level code, creating a disconnect between mental models and formal structures [75]. Vibe coding, by contrast, enables intent expression in natural language, aligning cognitive representation and emerging solutions while reducing extraneous cognitive load [75], [76], [77]. This process establishes closer coupling between the developer's mental model and the evolving artifact, reflecting design principles that favor matched mental models and immediate feedback [78], [79]. The alignment task shifts from API memorization to orchestrating a shared workflow that narrows the gap between intent and code structure, making cognition fundamentally interactive [80].

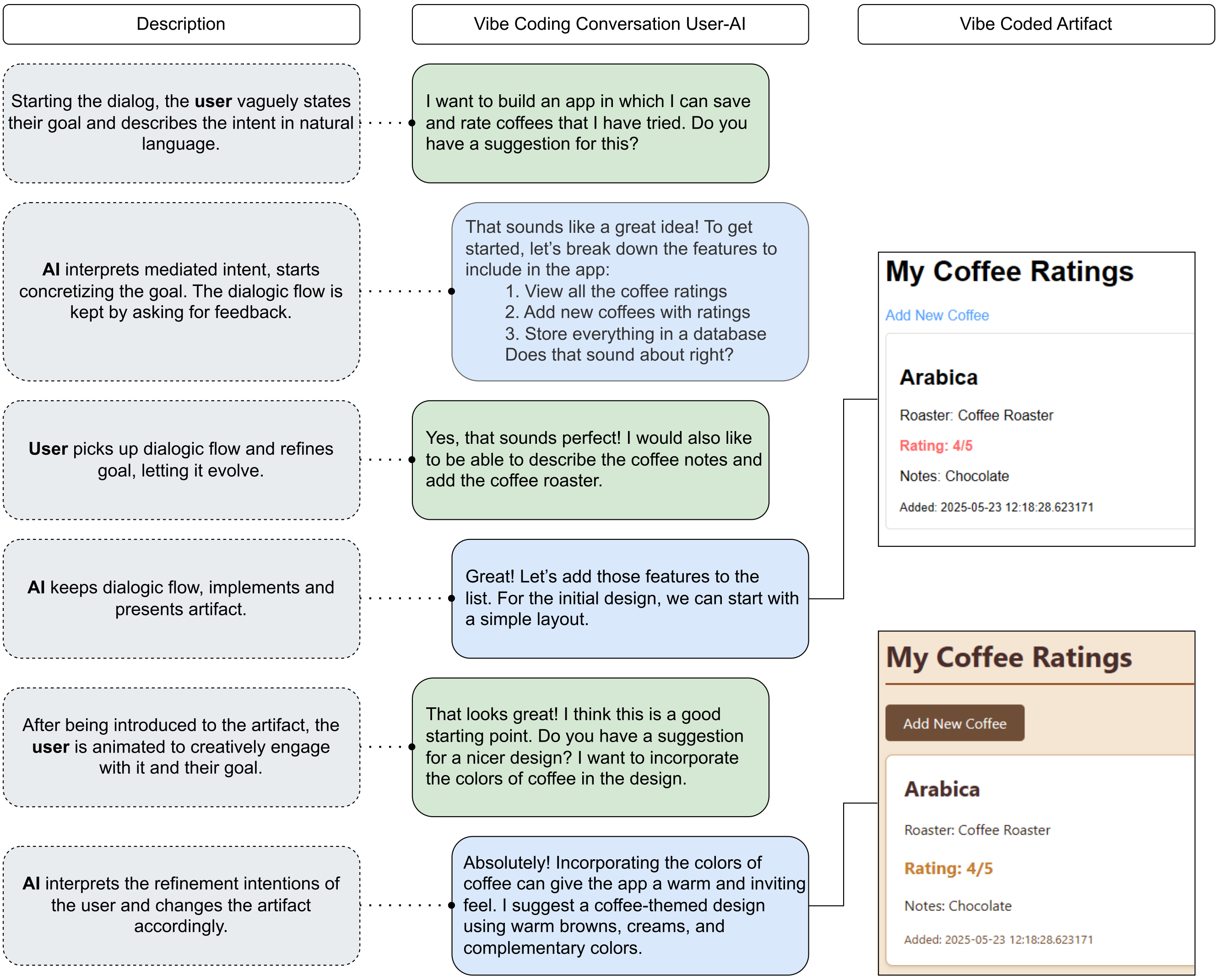

**FIGURE 1.** Dialogue between human user (green) and AI (blue) illustrating a vibe coding process.

Vibe coding transforms cognitive work from siloed individual burden to conversational co-creation, exemplifying distributed cognition across human and machine agents [14]. Developers articulate intentions in natural language, expressing goals and concepts while AI handles implementation, freeing cognitive resources for architectural decisions and problem conceptualization. By combining human creativity and strategic thinking with AI's recall and pattern implementation, the complementary strengths of both agents form a more efficient cognitive system than either could achieve alone, a concept referred to as "Hybrid Intelligence" [15]. Also, we argue that Vibe coding creates co-creative flow states with alternating leadership. Developers set goals and interpret results rather than writing every line of code. This resembles cognitive apprenticeship [81], where system feedback partially reveals the AI's "thought process" and enables collaborative learning. Cognitive roles thus shift from a single expert to a human-AI team, with humans as 'vibe directors' and semantic curators while AI serves as a dynamic problem-solver.

This redistribution alters temporal patterns of development as human-AI interactions become critical [82]. While traditional development imposed discrete write-compile-test-debug cycles fragmenting attention, vibe coding creates fluid, conversational cadence enabling faster feedback through dialogue-based iteration during ideation and design phases. While compile and execution cycles remain necessary for many applications, the conversational nature of the interaction can reduce cognitive fragmentation. Developers can make conceptual pivots more fluidly, keeping their focus on problem-solving rather than implementation details [83]. This iterative engagement aligns with agile principles [84], while vibe coding's emergent semantic alignment and acceptance of partial comprehension allow work to progress even when details aren't fully specified. Cognitive work becomes mutually adaptive, blurring the boundary between ideation and implementation.

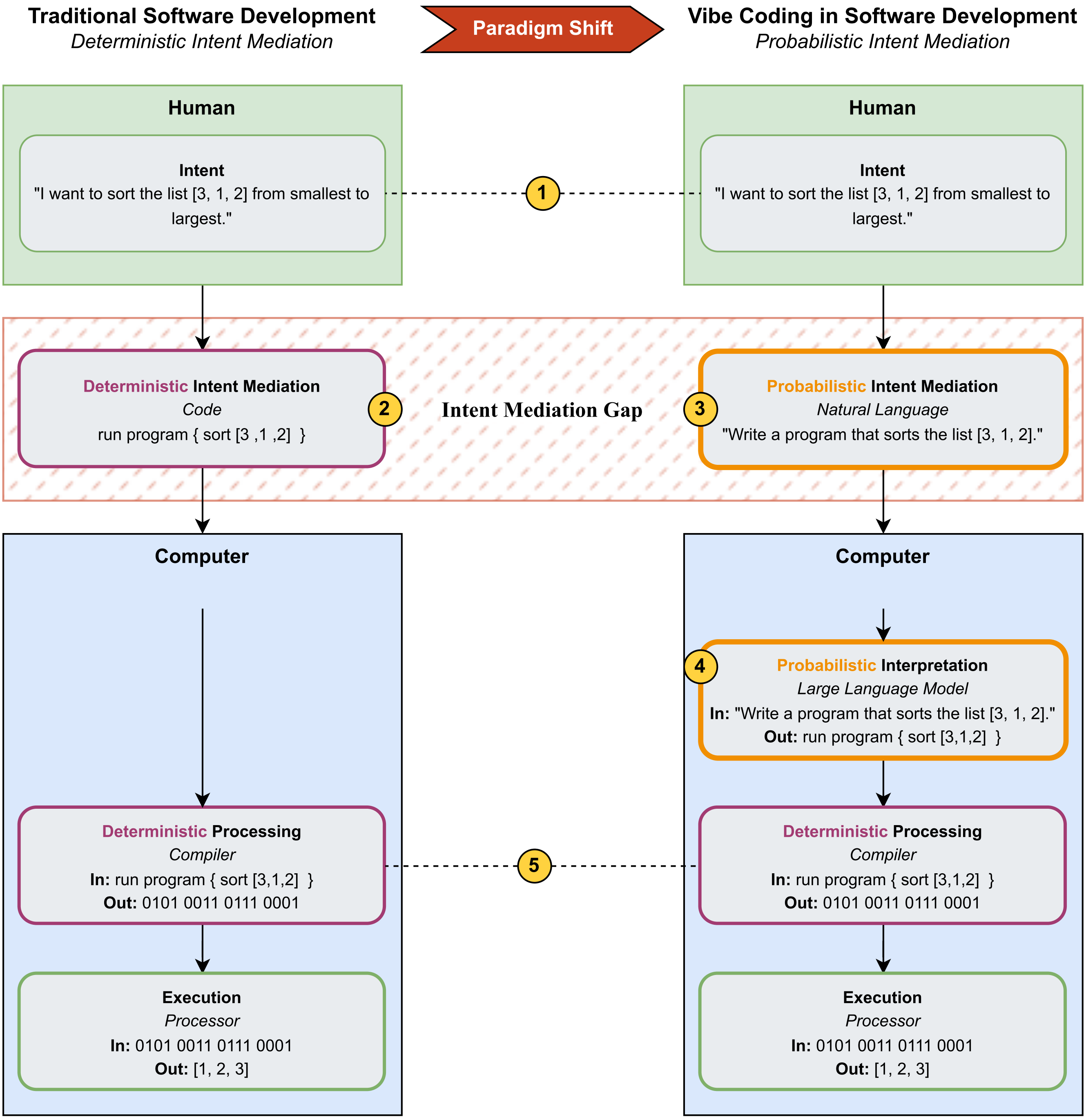

**FIGURE 2.** **Paradigm shift from deterministic to probabilistic intent mediation in software development.**

Traditional expertise is largely procedural: experts internalize idioms, mentally simulate execution, and excel in domain decomposition [85], [86], [87]. With AI assistance, tacit knowledge shifts to the tool. Vibe coding redefines expertise as adaptive collaboration, valuing problem framing, output validation, and design thinking. The expert becomes an orchestrator of vibes, steering co-creation. They elicit appropriate AI behavior and embrace evolving code. This connects to Polanyi's tacit knowing: developers "know more than they can tell" while AI generates code [86]. Expertise becomes metacognitive: asking right questions and interpreting answers. This favors technical knowledge blended with conversational skill. In vibe-oriented flow, developers navigate solutions through intuition, resembling jazz improvisation [83], [86]. Development expertise shifts from craftsmanship to synergistic problem-solving. The artifacts talk-back of the artifacts become more instant [79].

Traditional programming required comprehensive expertise in syntax, algorithms, and frameworks developed through years of practice [88]. Vibe coding transforms these boundaries by creating conversational ecosystems where AI compensates for human knowledge gaps, offsetting limited technical understanding [87]. This complementary relationship allows novices to produce functional code despite knowledge deficits, as AI interprets intentions and reduces cognitive load through natural dialogue rather than formal specification. The boundary in vibe coding changes from non-negotiable technical mastery to a permeable threshold where even partial understanding becomes acceptable. While Ericsson et al. emphasize how deliberate practice builds expertise within fixed domains [85], vibe coding expertise involves effectively guiding AI tools across flexible boundaries. The boundary's focus moves futher and further away from deep syntax knowledge toward problem decomposition, domain reasoning, and prompt engineering skills. Developers operating within the vibe coding paradigm still benefit from domain knowledge but face less rigid boundaries around language mastery. Critical thinking and effective AI collaboration become more crucial than memorizing syntactic details, as the complementary AI system handles boilerplate logic and compensates for technical knowledge gaps.

Finally, as knowledge boundaries become more permeable and expertise redistributes, questions of agency and responsibility emerge as critical considerations. Hence, aspects of epistemic agency and responsibility become central as vibe coding transforms knowledge ownership in software development. While traditionally developers were sole epistemic agents, vibe coding creates shared knowledge production. This resulting diffusion of authorship creates potential responsibility gaps [89]: when bugs arise, is the developer or tool at fault? The relational nature of agency in human-AI systems [90] complicates traditional notions of responsibility, as agency emerges from the interaction rather than residing in either the human or AI alone. Human developers retain ultimate agency but work with AI. Studies show humans tend to either over-rely on or under-utilize automated aids [91]. Despite accepting partial comprehension, developers must maintain skepticism by questioning AI suggestions. As Naeem & Hauser [92] note, users can integrate AI while maintaining responsibility. The dialogue-based workflow supports this by making AI "reasoning" partially explicit and editable. While knowledge creation becomes shared, developers transition toward orchestrator roles, remaining gatekeepers responsible for ensuring correctness, security, and goal alignment [89], [91].

The transformations in epistemic agency and responsibility highlighted above represent the culmination of vibe coding's reconfiguration of software development practice across multiple cognitive dimensions. These shifts, from sole to shared knowledge production, from complete comprehension to accepted partial understanding, and from individual to distributed responsibility, fundamentally alter how developers engage with code creation. Table 2 synthesizes these cognitive reconfigurations alongside the intent mediation transformations explored in Section III-A, providing a systematic comparison between traditional software development and the emerging vibe coding paradigm.

## IV. OPPORTUNITIES AND RISKS

As our analysis illustrates, vibe coding represents a paradigmatic change in how humans interact with systems to create software, fundamentally reconfiguring the relationship between human intent and machine execution through probabilistic rather than deterministic mediation. This transformation extends beyond productivity gains, reshaping who can develop, how programming tasks are approached, and the organizational structures that result. Drawing on the dimensions of intent mediation, cognitive work, and expertise outlined in Table 2, we further theorize about vibe coding's implications, positing that this evolution presents significant opportunities as well as risks that reflect a complex interplay that likely only hints at the broader consequences.

### *A. OPPORTUNITIES OF VIBE CODING*

The shift toward probabilistic intent mediation in software development creates opportunities ranging from individual empowerment to organizational transformation. By democratizing development, enabling cognitive liberation, accelerating feedback loops, and providing systemic leverage, vibe coding changes not only how software artifacts are produced but also who can participate in digital creation. These opportunities span multiple dimensions of development practice, suggesting that the conversational paradigm will profoundly impact both the social and technical fabric of software development, with many consequences still emerging.

#### 1) COGNITIVE ACCESSIBILITY AND INCLUSION

Natural language and multimodal interfaces are transforming software development from specialized craft to expressive medium accessible to broader, more diverse participants.

**TABLE 2. Traditional Software Development vs. Vibe Coding.**

| Category | Dimension | Traditional Software Development | Vibe Coding |
|---|---|---|---|
| Intent Mediation | Intent Layers | Intent expressed in formal programming languages; limited to text mode via code editors. | Intent expressed through natural language, voice, visual cues; supports multimodal interaction across chat, speech, or graphical interfaces. |
| | Intent Translation | Human decomposes intent into semantics and syntax; compiler handles syntax-to-execution. | AI infers semantics from naturalistic input and generates syntax; developer guides through prompt iteration and review. |
| | Intent Fidelity | High fidelity through explicit specification and manual control. | Variable fidelity; AI interpretation introduces ambiguity, requiring testing and refinement. |
| Cognitive Work | Cognitive Alignment | Requires structured, abstract reasoning aligned with machine logic and formal languages. | Aligns with intuitive, expressive reasoning; supports informal articulation of goals. |
| | Configuration of Cognitive Work | Developer as sole constructor and debugger; responsible for all formalization. | Developer as articulator, critic, and tester; shares generative and interpretive labor with AI. |
| | Cognitive Rhythm of Engagement | Linear and staged: plan → implement → test; feedback is delayed and tool-mediated. | Dialogic and iterative: prompt → interpret → revise in near real time; continuous co-adaptation. |
| Expertise | Nature of Expertise | Emphasizes formal implementation and optimization. | Emphasizes articulation, prompting, validation, and strategic steering of generative systems. |
| | Knowledge Boundary | Fixed and comprehensive; requires mastery of syntax, algorithms, architecture, and system interaction with minimal external compensation. | Fluid and complementary; AI capabilities offset human knowledge gaps through iterative dialogue, reducing need for technical mastery. |
| | Epistemic Agency and Responsibility | Human holds full authorship and explanatory authority over code behavior. | Epistemic agency is shared; AI proposes logic, human accepts, tests, and assumes partial accountability. |

By enabling communication through everyday language and intuitive interactions, these paradigms lower barriers to entry and foster inclusive innovation. This reconfiguration of the landscape opens several pathways for inclusion.

By enabling AI to infer intent from natural language, domain specialists can directly translate their expertise into functional software. This *empowerment of domain expertise* reduces the reliance on formal programming skills and rebalances traditional technical hierarchies [93].

A *lowered entry threshold* allows individuals without technical backgrounds to participate meaningfully in software development by programming that aligns with intuitive reasoning and thus broadening access and democratizing technological creation [94].

AI's ability to complement human knowledge enables users to make progress even with incomplete technical understanding. This *cognitive scaffolding* supports productive work without requiring full mastery of implementation details [95].

### 2) COGNITIVE LIBERATION

Vibe coding reframes development from technical execution to strategic orchestration through collaborative dialogue that shapes both problem and solution. This privileges iterative refinement and higher-order thinking over predetermined steps. As a result, vibe coding supports a more reflective and adaptive form of problem-solving, where knowledge is constructed collectively and solutions emerge through interaction and negotiation.

With AI's ability to infer intent from naturalistic input, *rapid prototyping and iteration* is made possible. This accelerates the creation and exploration of initial solutions, enabling rapid testing and refinement of alternative approaches before committing to a final implementation [96], [97].

Enabling *Human-AI Co-Creation* by shifting cognitive work from individuals to collaborative human-AI configurations, allows developers to act as articulators, critics and testers. This partnership frees human resources for strategic and creative tasks while AI manages implementation complexity [98].

As programming emphasizes articulation, prompting, and strategic steering over implementation, *new forms of expertise* emerge. Evaluative and dialogic skills gain prominence over low-level mastery. This expertise evolution creates space for differently skilled individuals to contribute meaningfully to software development [99].

Within the dialogic and iterative development rhythm, requirements evolve through AI interaction rather than upfront specification, thus necessitating *emergent problem understanding*. This emergent approach encourages exploratory design thinking and reduces cognitive overhead in early development stages [99].

By *expanding creative horizons*, freeing from the constraints of formal syntax and implementations, developers can Within the dialogic and iterative development rhythm, requirements evolve through AI interaction rather than upfront specification [100].

#### 3) ACCELERATED DEVELOPMENT CYCLES

Traditional software development typically progresses through linear stages such as requirements gathering, design, implementation, testing, and deployment. These handoffs often create bottlenecks and limit opportunities for feedback and revision. In contrast, the continuous co-adaptation inherent in vibe coding enables more fluid approaches to system construction.

AI's ability to infer intent from naturalistic input accelerates the creation and exploration of initial solutions, enabling *rapid prototyping, iteration*, and refinement of alternative approaches before committing to a final implementation. This revolutionary speed underscores the core concept of Vibe Coding [101].

By automating the translation from abstract requirements to executable code, the *translation overhead* is reduced. Thus, AI minimizes manual mapping and cognitive load for engineers, expediting development and reducing error rates [102].

The experience is enhanced by a *conversational and iterative flow*. Real-time, dialogic engagement with AI fosters a seamless development rhythm, promoting a creative flow and minimizing friction between ideation and realization [103].

The process of vibe coding allows for *flexible, evolving specifications*. Development proceeds through cycles of prompting, interpretation, and revision, enabling solutions to emerge organically without the need for fully specified requirements at the outset [104].

#### 4) SYSTEMIC LEVERAGE

Beyond its advantages for individuals, vibe coding generates transformational impacts at both the organizational and ecosystem levels. These changes help organizations become more agile and responsive to emerging challenges and opportunities. In this way, the opportunities previously identified, such as increased inclusion, enhanced problem-solving, and more flexible system development, are elevated from isolated benefits to strategic outcomes that shape the direction and success of entire organizations and broader communities.

The adoption of AI and conversational workflows fundamentally restructures organizational capabilities, such as *scalable team efficiency*. By amplifying the output of small teams, this approach enables organizations to achieve results that previously required much larger technical departments. This democratizes innovation, empowering startups and smaller units to compete effectively with established industry players [105].

This shift also facilitates a strategic *talent reallocation*. As the nature of cognitive work transforms from implementation to orchestration, organizations can prioritize hiring domain experts and creative thinkers over traditional technical specialists. This optimizes talent deployment and fosters richer interdisciplinary collaboration across the board [106].

Accessible interfaces and conversational workflows allow organizations to pursue projects previously constrained by significant technical complexity or resource limitations, thus broadening the scope of feasible innovation [107].

The synergy of rapid prototyping, reduced translation overhead, and iterative refinement compresses development cycles, leading to accelerated innovation and market responsiveness. This alignment between emergent solutions and market needs enables faster, more relevant innovation, which ultimately provides a stronger competitive edge [108].

### B. RISKS OF VIBE CODING

Despite its promising opportunities, vibe coding introduces significant challenges that warrant critical examination as the paradigm shift toward probabilistic intent mediation reshapes software development practice. The risks span from potential erosion of technical expertise and degradation of code quality to the emergence of responsibility gaps and organizational vulnerabilities that threaten long-term sustainability. Real-world incidents emerging from early vibe coding adoption illustrate that these risks often manifest in areas like security, reliability, and autonomy. Addressing these challenges requires proactive strategies for knowledge preservation, quality assurance, strategic planning and governance, which encompasses policies and structures that regulate AI tool use, establish oversight requirements, and assign responsibility for outcomes. Only by anticipating and managing these risks can we ensure that the benefits of vibe coding are realized without compromising fundamental software engineering principles or introducing unforeseen negative consequences.

#### 1) EROSION OF PROGRAMMING EXPERTISE

As AI increasingly mediates programming through naturalistic input, traditional expertise in code manipulation, syntax understanding, and procedural application may become less central, raising questions about how expertise is now formed, maintained, and transferred.

With developers acting more as articulators than implementers, core technical skills, such as algorithmic thinking, debugging, and architectural planning, may atrophy. Studies on Generative AI capabilities suggest that this *deskilling* or "leveling of ability" is a common outcome [109]. Reliance on AI to fill knowledge gaps can consequently weaken a developer's capacity for systematic solution design and independent problem-solving.

The ability to generate functional code and complex artifacts without requiring deep domain understanding fosters a *false sense of competence*, increasing the risk of overlooking critical issues like security and performance. Illustrating security-related risks, research indicates that participants who had access to an AI assistant were more likely to believe they wrote secure code, suggesting that such tools may lead users to be overconfident about security flaws [110], [111].

As development evolves increasingly toward prompt engineering, traditional and essential avenues for *transmitting tacit knowledge*, such as direct mentorship, pair programming, and rigorous code reviews, are diminished. This impedes the professional growth of less experienced developers, an issue that requires particular attention to be given to professional development and curriculum reform [112]. The long-term negative impact on novice skill acquisition further reinforces this threat [113].

#### 2) CODE QUALITY AND MAINTAINABILITY

Generative AI integration risks undermining collective practices ensuring code stability. Teams rely on shared standards, review processes, and systematic testing to maintain quality and identify errors. As AI produces more code, these collaborative routines may weaken.

Iterative prompting can lead to unpredictable code rewrites, which makes local debugging less effective and introduces *instability* with unanticipated side effects. This prioritization of immediate functionality over sound architecture accelerates technical debt and creates long-term maintenance burdens, a common issue as AI-based systems manage and accrue complexity [114]. The discovery of security flaws in commercial vibe coding platforms underscores security and autonomy risks in practice [115].

Additionally, developers face the problem of *opaque verification and inconsistent patterns*. As traditional reviews and testing protocols are replaced or influenced by AI-driven judgments, quality assurance becomes less transparent and reliable. Highlighting reliability concerns, large-scale security analysis of over 100 language models revealed that only 55 % of AI-generated code was secure, with security performance remaining largely flat over time even as models improved at generating syntactically correct code [116]. This suggests that functional correctness does not guarantee security quality in AI-generated code. Furthermore, code generated through conversational flows often lacks cohesive structure, consistent patterns, and proper documentation, greatly complicating future maintenance. Reflecting reliability challenges, research highlights that AI-generated code, while simpler, is more prone to unused constructs and hardcoded debugging, which contrasts with human-written code that, despite having greater structural complexity, generally exhibits higher maintainability [117].

#### 3) EPISTEMIC AND RESPONSIBILITY GAPS

Shared epistemic agency distances developers from underlying logic and intent. Distributed responsibility makes it difficult to understand, explain, or assign accountability for outcomes when problems arise, complicating ethical oversight.

A salient concern pertains to the *ambiguity surrounding authorship and accountability*. Dialogic, iterative engagement blurs the boundaries of ownership between human and AI, making it difficult to assign responsibility for errors or unethical outcomes, thereby creating significant governance challenges [118], [119].

As developers increasingly rely on AI-generated code, they become detached from the internal logic of their systems. This partial comprehension makes it harder to understand, intervene, or recover during critical failures, particularly when deep system knowledge is essential. The challenges facing AI-generated code include accuracy, contextual understanding, security, privacy, and ethical considerations, necessitating thorough review and testing [120]. This challenge is specifically known as the *AI Black Box Effect* [121].

the continuous co-adaptation workflow leads to the *loss of intent traceability*. This obscurity makes it difficult to connect original requirements to implemented solutions. Without transparent mapping between prompts, revisions, and the final code, future maintainers cannot reconstruct the rationale behind key decisions, even though traceability is known to improve software maintenance quality [122].

The inherent limitations of the technology create *ethical and data protection blind spots*. AI's limited explanation capabilities and reliance on probabilistic mediation increase the risk of undetected ethical breaches and data protection violations. These issues may only become apparent in production environments or edge cases, necessitating careful scrutiny of AI's output regarding security, privacy, and ethical compliance [120].

#### 4) STRATEGIC AND ORGANIZATIONAL VULNERABILITIES

Vibe coding introduces systemic challenges affecting organizational structures and ecosystems. The collaborative nature complicates maintaining consistent processes, enforcing standards, and ensuring clear responsibility, while interconnectedness can amplify problems across teams and organizations.

The convergence of black box codebases, ethical and data protection blind spots, and inconsistent documentation severely undermines a system's auditability and verifiability. This poses significant barriers to compliance in highly regulated sectors such as healthcare and finance. Expert consensus widely acknowledges the need to establish clear responsibility and accountability for the outputs and impacts of AI-enabled systems, making transparency essential for justice and governance [123].

The current AI landscape introduces significant ecosystem bias and tool dependency. AI inference capabilities are stronger for mainstream programming languages, which grants a strategic advantage to some organizations while disadvantaging those using specialized languages. Simultaneously, reliance on proprietary AI tools and their probabilistic mediation introduces economic and infrastructural dependencies. This reduces organizational autonomy and exposes companies to risks like pricing changes, service discontinuations, or shifting external priorities. These factors, described as a "Matthew Effect," reinforce existing popularity hierarchies among tools, creating competitive distortions and long-term strategic vulnerabilities [124].

As organizational expertise shifts toward prompt engineering and strategic steering, traditional software development skills, such as authoring code, debugging, maintenance, and scaling, risk being undervalued. This misalignment threatens system continuity, particularly when deep technical knowledge is required for critical systems. The future role of software engineers will continue to change, requiring adaptation of the profession to remain relevant and effective in an AI-assisted environment [125].

## V. DISCUSSION

### A. VIBE CODING AND THE RECONFIGURATION OF INTENT MEDIATION

Vibe coding represents a shift from deterministic to interpretive and collaborative development.

The nature of software development expertise is hence fundamentally transformed. As discussed in Section III, traditional skills such as implementation-specific fluency and syntactic mastery are increasingly supplanted by new competencies, including problem articulation, prompt engineering, and evaluative judgment. Research on live programming environments highlights the importance of immediate feedback on AI-generated code in fostering these emerging skills [127]. Moreover, recent studies emphasize that the widespread adoption of large language models in software engineering amplifies these challenges, raising critical concerns about code quality, explainability, and the urgent need for updated educational and professional practices [128]. This evolution reflects a redistribution of expertise across a collaborative human-AI system.

As outlined in Section III and Section IV, vibe coding presents opportunities for democratization, acceleration, and enhanced cognitive accessibility in software development. However, it also introduces risks, including technical deskilling, code quality issues, and responsibility gaps. Empirical studies indicate that programming with large language models can yield significant productivity gains [17], but may also introduce new forms of systemic fragility. Recent analysis identifies significant challenges inherent to LLM-assisted code generation, including insecure code generation, hallucinated outputs, irreversible actions, vulnerability inheritance, overtrust, and the absence of standardized validation and rollback protocols [111]. These concerns are compounded by the accumulation of technical debt, as AI-generated code often lacks the structure, documentation, and clarity necessary for long-term maintenance, leading to increased costs and making future modifications and debugging significantly more difficult [126]. The discovery of a critical vulnerability in Wix's Base44 vibe coding platform, which allowed unauthorized access to private enterprise applications through simple exploitation techniques, underscores security and autonomy risks in practice, with researchers warning that fast-paced vibe coding platforms may introduce systemic risks to entire ecosystems [115]. Addressing these challenges will require innovative approaches to software development, as well as revised educational and professional practices. For instance, research has shown that the quality of identifier construction directly affects a developer's ability to comprehend and debug code, suggesting that AI-generated code with ambiguous or poorly chosen names may exacerbate these difficulties [129].

Current institutional frameworks, which rely on explicit control and procedural transparency, are destabilized by vibe coding's shared epistemic agency and fluid knowledge boundaries. This misalignment calls for coordinated adaptation across educational, regulatory, and organizational domains to establish coherent frameworks for this new programming paradigm. Furthermore, the digital environments in which software development now occurs may subtly nudge practitioners toward certain approaches, creating choice architectures that institutional frameworks have yet to recognize or address. Regulatory systems, which assume clear lines of accountability, are challenged by the blurred responsibilities inherent in collaborative human-AI development. The opacity of AI decision-making processes further complicates accountability, making explainable Artificial Intelligence essential for maintaining institutional oversight and compliance.

### B. FUTURE RESEARCH DIRECTIONS

The shift to probabilistic intent mediation reveals gaps in understanding how Generative AI reconfigures development practice, expertise, and organizational structures. While offering opportunities for democratization and acceleration, vibe coding introduces uncertainties requiring systematic investigation and governance frameworks. The urgency of establishing such frameworks is underscored by rapid AI adoption in business contexts, where organizations must navigate complex governance challenges. Drawing from a sociotechnical perspective, we organize this research agenda around three interconnected domains: the human actors whose cognitive work and professional identities are being reconfigured, the technological systems whose interpretive capabilities shape what can be expressed and built, and the organizational structures that must evolve to accommodate shared epistemic agency.

#### 1) HUMAN-CENTERED RESEARCH DIRECTIONS

Vibe coding transforms how developers think, learn, and define themselves, necessitating rethinking cognitive models, expertise, and pedagogical frameworks while opening possibilities for broader inclusion. The current professional development has a potential to swap over to the field of End-User-Development. As natural language becomes a primary interface to create artifacts, traditional forms of mastery may be replaced, or complemented, by new types of articulation and collaborative fluency.

1) How can the new environment be used for software development education and what are appropriate didactical approaches for teaching software development?
2) How must software development education be restructured to develop competencies in prompt articulation, semantic validation, and co-creative orchestration?
3) What forms of expertise are emerging in vibe coding, and how should these be systematically recognized, assessed, and certified?
4) How does vibe coding reconfigure the identity of developers, and what are the implications for professional development and recruitment?
5) Does vibe coding measurably lower participation thresholds for underrepresented groups in software development, and how can this potential be institutionalized?

#### 2) TECHNOLOGY-CENTERED RESEARCH DIRECTIONS

Vibe coding's probabilistic processes remain opaque, raising questions around intent fidelity, semantic alignment, and explainability. While multimodal prompts allow flexibility, they introduce ambiguity and variability complicating reproducibility. Research must focus on transparency, accountability, and certifiability for legal, ethical, or safety-critical workflows.

1) How do different LLMs vary in their interpretation of identical prompts, and what are the systemic implications for reliability, reproducibility, and control?
2) What linguistic and modal features ensure high-fidelity intent expression across LLMs, and can these be codified into standardized prompt design guidelines?
3) How can explainability be operationalized in vibe coding environments to render AI inferences interpretable and critique-ready?
4) What technical and procedural standards are required to certify AI-generated code for use in critical domains with legal, ethical, or safety implications?

#### 3) ORGANIZATION-CENTERED RESEARCH DIRECTIONS

Vibe coding reshapes institutional logic, team composition, and accountability structures. Organizations must adapt methodologies, redefine roles, and implement governance protocols for shared human-AI authorship while maintaining critical technical capabilities. Strategic alignment, compliance, and knowledge retention emerge as key concerns.

1) What new roles must organizations establish to manage prompt engineering, semantic design, and AI oversight in software teams?
2) How should software development processes be restructured to support dialogic iteration, emergent requirements, and epistemic ambiguity?
3) How can accountability and legal liability be operationalized in systems where authorship and agency are shared between humans and AI?
4) How can organizations preserve deep programming expertise for system resilience while scaling vibe coding practices operationally?

These research directions reflect the complex reconfiguration of software development as both a technical and social practice. Vibe coding challenges longstanding assumptions about intent, authorship, expertise, and responsibility in software creation. Addressing these challenges will require interdisciplinary collaboration across computing, design, education, organizational science, and ethics. In particular, ethical inquiry should engage with how vibe coding inherits and potentially amplifies concerns already central to AI ethics. Biases latent in training data may propagate undetected into deployed systems, while the layered opacity of AI-generated code, where neither model nor developer may fully comprehend the implementation, complicates meaningful auditing for fairness, safety, or discriminatory effects. Moreover, as software creation becomes increasingly dependent on proprietary AI systems controlled by few corporations, vibe coding raises broader questions about epistemic dependence, the erosion of developer capacity to scrutinize and challenge generated outputs, and the concentration of power over the tools that shape how software is made. A robust socio-technical research agenda, attending equally to technical architectures and to the human conditions, values, and vulnerabilities they encode, is therefore essential not only for understanding vibe coding, but for shaping its institutional, technical, and cognitive future.

## VI. CONCLUSION

This paper defines vibe coding as a software development paradigm characterized by natural language dialogue and collaborative flow between humans and AI. Situating it within the historical evolution of intent mediation highlights how shifts in modality reconfigure the human-machine relationship. Addressing the first research question, the paper has articulated how vibe coding distinguishes itself from traditional software development by shifting intent mediation from explicit instruction to probabilistic, goal-oriented dialogue, as reflected in its five key attributes: goal-oriented intent expression, rapid dialogic interaction, implementation abstraction, dynamic semantic refinement, and co-creative flow states.

In response to the second research question, the analysis has explored the cognitive, epistemic, and organizational implications of vibe coding. The opportunities identified include enhanced cognitive accessibility and inclusion, cognitive liberation, accelerated development cycles, and systemic leverage. However, these benefits are accompanied by significant risks, such as the erosion of programming expertise, challenges to code quality and maintainability, epistemic and responsibility gaps, and strategic and organizational vulnerabilities. While the definition and implications outlined in this paper offer a foundation for understanding vibe coding, they remain open to refinement as the field develops.

Looking ahead, the future research questions proposed here point toward human-centered, technology-centered, and organization-centered research directions. Continued investigation along these lines will be essential for establishing a nuanced understanding of vibe coding, enabling the field to harness its benefits while addressing its inherent risks as this development modality matures.

## ACKNOWLEDGMENT

The authors used ChatGPT-4o (OpenAI) and Claude Sonnet 4 (Anthropic) across all sections of the manuscript for language-related support, including proofreading, rephrasing, structuring, or editing of their-drafted text. All content was reviewed by them, who take full responsibility for the final version of the manuscript.

**CHRISTIAN MESKE** is a Full Professor of socio-technical system design and artificial intelligence with the Faculty of Mechanical Engineering and the Institute of Work Science, Ruhr University Bochum, Germany. He has contributed to numerous interdisciplinary research projects and published more than 80 articles in leading conferences and journals, including *Applied Energy*, *Business and Information Systems Engineering*, *Communications of the Association for Information Systems*, *Information Systems Journal*, *Information Systems Management*, *Journal of the Association for Information Systems*, *MIS Quarterly Executive*, and *Solar Energy*. He also serves as a Senior Editor for *Information Systems Management*, an Associate Editor for *Behaviour and Information Technology*, and holds various other roles.

**TOBIAS HERMANNS** received the M.Sc. degree in applied computer science from Ruhr University Bochum, Germany. He is currently pursuing the Ph.D. degree with the Chair of Socio-technical System Design and Artificial Intelligence, Faculty of Mechanical Engineering and the Institute of Work Science, and affiliated with the Faculty of Computer Science with Ruhr University Bochum. In parallel to his studies, he worked at the Chair of Embedded Security in collaboration with the Max Planck Institute for Security and Privacy, contributing to the development and maintenance of "HAL–The Hardware Analyzer", an open-source tool for reverse engineering hardware, used in both research and teaching. His research focuses on human–computer interaction and user experience in the context of generative artificial intelligence, especially in relation to interaction design with large language models and the dynamics of human–AI collaboration. His interests include human–computer interaction, human–AI interaction, Generative AI, and socio-technical system design.

**ESTHER VON DER WEIDEN** received the M.Sc. degree in applied computer science from the University of Duisburg–Essen, Duisburg, Germany. She is currently pursuing the Ph.D. degree with the Chair of Socio-Technical System Design and Artificial Intelligence, Data Protection Office, Ruhr University Bochum, Bochum, Germany. She is currently a Research Associate with the Chair of Socio-Technical System Design and Artificial Intelligence, Data Protection Office, Ruhr University Bochum. Her research focuses on the intersection of data protection and artificial intelligence. Her research interests include data protection, generative AI, information engineering, AI, and information literacy.

**KAI-UWE LOSER** received the Ph.D. degree in computer science from Technical University Dortmund, Dortmund, Germany. He is also a Certified Data Protection Auditor. He serves as the Official Data Protection Officer with Ruhr University Bochum and the University Duisburg–Essen, and is Vice President of the Professional Association of Data Protection Officers (BvD). He has several publications in peer-reviewed journals and international conference proceedings. His research interests include data protection, modeling of socio-technical systems, knowledge management and learning organizations, and the application of groupware to support knowledge management and learning organizations.

**THORSTEN BERGER** is currently a Professor in computer science with Ruhr University Bochum, Germany. He published more than 180 papers, many in A* venues, such as ICSE; FSE, and ASE, and Q1 journals such as IEEE TSE. He is the co-author of the textbook *Domain-Specific Languages: Effective Modeling, Automation, and Reuse*. His research focuses on AI engineering, software variability, model-driven engineering, and software security. He received competitive grants from Swedish Research Council, Wallenberg Autonomous Systems Program, Vinnova Sweden (EU ITEA), and European Union, as well as he was awarded the Fellowship of the Royal Swedish Academy of Sciences and of the Wallenberg Foundation—one of the highest recognitions for researchers in Sweden. He received three most-influential-paper, two best-paper, and three distinguished-reviewer awards.